\documentclass[aps,twocolumn]{revtex4}

\begin{document}

\bibliographystyle{apsrev}

\title{Gravitational Wave Confusion Noise}
\author{ Neil J. Cornish}
\affiliation{Department of Physics, Montana State University, Bozeman, MT 59717}

\begin{abstract}
One of the greatest challenges facing gravitational wave astronomy in the low frequency
band is the confusion noise generated by the vast numbers of unresolved galactic and
extra galactic binary systems. Estimates of the binary confusion noise suffer from
several sources of astrophysical uncertainty, such as the form of the initial mass function
and the star formation rate. There is also considerable uncertainty about what defines
the confusion limit. Various ad-hoc rules have been proposed, such as the one source
per bin rule, and the one source per three bin rule. Here information theoretic methods are
used to derive a more realistic estimate for the confusion limit. It is found that the
gravitational wave background becomes unresolvable when there is, on average, more than one
source per eight frequency bins. This raises the best estimate for the frequency at which
galactic binaries become a source of noise from 1.45 mHz to 2.54 mHz.
\end{abstract}

\maketitle

Population synthesis models and astronomical observations suggest that there may be as many
as $2 \times 10^8$ galactic disk binaries containing two compact objects. Several million of
these binaries are expected to emit gravitational waves that can be detected by the
Laser Interferometer Space Antenna (LISA). At frequencies below $\sim 1$ mHz these systems
constitute an embarrassment of riches for gravitational wave astronomy as the number of
sources is so great that it becomes impossible to isolated individual binaries. The unresolved
galactic gravitational wave background is a source of noise that degrades
LISA's ability to detect other sources, such as extra-galactic supermassive black hole binaries.

A much quoted definition of the gravitational wave confusion noise level is the amplitude at
which there is, on average, at least one source per frequency bin. Even this
crude definition takes into account some data analysis considerations as the width of a frequency
bin, $\Delta f = 1/T$, depends on the observation time $T$. The longer the observation, the
smaller the confusion noise. However, Hellings \cite{ron} pointed out that there was not
enough information in one frequency bin to fully specify a binary system. A monochromatic,
circular binary is described by the seven parameters $(f,\ln {\cal A}, \theta, \phi,
\imath, \psi, \varphi_0)$, where $f$ is the gravitational wave frequency,
${\cal A}$ is the intrinsic amplitude, $(\theta,\phi)$ describe the sky location,
$\imath$ denotes the inclination, $\psi$ is the polarization angle and $\varphi_0$ is
the orbital phase. Hellings reasoned that each frequency bin only contains
two pieces of information - the real and imaginary parts of the strain spectral density -
while six pieces of information are needed to fix $(\ln {\cal A}, \theta,
\phi, \imath, \psi, \varphi_0)$. Thus, at least three frequency bins are required to solve
for the six parameters. The frequency was not included in the information budget on the
grounds that it came for free from the Fourier transform.
While the three bin rule is an improvement over the naive one bin rule, it
lacks a rigorous justification. More recently, Phinney \cite{sterl} has revisited
the confusion noise question by using Shannon information theory to estimate
the total number of binaries than can be resolved by LISA. Here a similar approach is used to
estimate the number of frequency bins that are needed to resolve a binary system.

The estimate is made by comparing the information contained in one frequency bin with
the amount of information required to record the parameters that describe a binary system.
Both of these quantities depend on the signal to noise ratio of the instrument.
A higher signal to noise ratio means more information in each bin, but it also
means that the binary parameters are better determined, and thus take more information
to describe.

The Baud rate of the detector, $C$, depends on the bandwidth of the signal, $B$,
and the signal to noise ratio $S/N$ according to
\begin{equation}
C = B \, {\rm log}_2 \left(1 + \frac{S}{N}\right) \, .
\end{equation}
The total number of bits transmitted during an observation time $T$ is simply $CT$. Setting
the bandwidth equal to $\Delta f=1/T$, it follows that the
amount of information per bin is equal to $\Delta I={\rm log}_2 \left(1 + \frac{S}{N}\right)$.
Because the LISA observatory is designed to operate in stereo, {\it i.e.} LISA can detect
both gravitational wave polarizations \cite{curt}, the total amount of information available
in each frequency bin is larger than $\Delta I$, but not a full factor of two larger as
the information in each channel is not fully independent. The analysis of the stereo performance
is further complicated by the correlated noise in the two channels. In what follows, the
information budget is calculated for a single channel, with the expectation that the result
would be little changed by including the second channel.

The amount of information required to record the parameters of a binary system can be
estimated by comparing the volume of the error ellipsoid to the volume
of the parameter space. To see how this works, consider the following
one dimensional example. Suppose that the angle $\phi$ can be measured to an accuracy of
three degrees. Recording $\phi$ to this accuracy takes $\log_2(360/3)\approx 7$
bits of data. The volume of the seven dimensional error ellipsoid that characterizes the
binary subtraction problem for LISA can be estimated by calculating the
determinant of the Fisher information matrix $\Gamma_{ij}$. To calculate
the Fisher matrix an expression is needed that describes how information about the
binary is encoded in the the detector output. At low frequencies, the signal in each LISA
channel takes the form
\begin{eqnarray}
s_(t) &=&  F^+(t,\psi,\theta,\phi)A_+({\cal A},\imath)
\cos\Psi(t,f,\theta,\phi,\varphi_0)
\nonumber \\
&& + F^\times(t,\psi,\theta,\phi)A_\times({\cal A},\imath)
\sin\Psi(t,f,\theta,\phi,\varphi_0)\, ,
\end{eqnarray}
where
\begin{equation}
\Psi(t,f,\theta,\phi,\varphi_0) = 2\pi f t+\Phi_D(t,f,\theta,\phi)+\varphi_0 \, .
\end{equation}
Expressions for the detector response functions $F^+$ and $F^\times$, the Doppler modulation
$\Phi_D$, and the gravitational wave amplitudes $A_+$ and $A_\times$ can be found in
Ref. \cite{cr1}. The components of the Fisher matrix are found using the expression \cite{curt}
\begin{equation}\label{fisher}
\Gamma_{ij}(\vec{\lambda}) =  \frac{1}{S_n(f)}
\int_0^T \partial_i s_\alpha(t) \partial_j s_\alpha(t) \, dt \, ,
\end{equation}
where the components $i,j$ range over the seven parameters $\vec{\lambda} \rightarrow
(f,\ln {\cal A}, \theta, \phi,
\imath, \psi, \varphi_0)$, and $S_n(f)$ is the one-sided noise spectral density. In general,
the Fisher matrix is a complicated function of all seven parameters, which implies that
the volume of the error ellipsoid varies depending on where the binary sits in
parameter space. Assuming that the noise is Gaussian, and working in the limit where the
signal is large compared to the noise, the volume of the seven dimensional error
ellipsoid is give by
\begin{equation}\label{vol}
\Delta V(\vec{\lambda})= \frac{\pi^3\sqrt{2}}{105} \sqrt{{\rm det}(C_{ij}(\vec{\lambda}))} \, ,
\end{equation}
where $C_{ij} \simeq \Gamma_{ij}^{-1}$ is the noise covariance matrix,
$C_{ij}=\langle \Delta \lambda_i \Delta \lambda_j \rangle$.
The amount of information needed to describe or
subtract a binary source emitting gravitational waves with frequency $f$ is given by
\begin{equation}\label{sinfo}
{\cal I}(B,\vec{\lambda}) = \log_2 \left(\frac{ V(B)}{\Delta V(\vec{\lambda})} \right) \, ,
\end{equation}
where $V(B)$ is the volume of the parameter space in a bandwidth $B$.
The bandwidth needed to subtract a typical binary can be estimated by solving the
transcendental equation
\begin{equation}\label{tran}
{\cal I}(B) = BT \log_2 \left(\frac{S}{N}\right) \, ,
\end{equation}
for $B$. Here ${\cal I}(B)$ is the average amount of information required to remove one binary,
and $S/N \gg 1$ is the average signal to noise ratio across the bandwidth. Note
that the left hand side of the transcendental equation depends logarithmically on the
bandwidth and the observation time, while the right hand side depends linearly on these
quantities. 

Calculating ${\cal I}(B,\vec{\lambda})$ is a straightforward, yet computationally intensive task. 
The volume of the error ellipsoid can be expressed in terms of the eigenvalues $\sigma^2_i$
of the covariance matrix $C_{ij}$:
\begin{equation}
\Delta V(\vec{\lambda})= \frac{\pi^3\sqrt{2}}{105} \prod_{i=1}^{7} \sigma_i(\vec{\lambda})\, .
\end{equation}
The eigenvectors of the covariance matrix lie along the principle axes of the
error ellipsoid, which may not be aligned with the coordinate axes. The
Fisher matrix can be expressed in the form
\begin{equation}\label{fish}
\Gamma_{ij}(\vec{\lambda}) = \left(\frac{S}{N}\right)^2 \gamma_{ij}(f,\theta,\phi, \imath, \psi)
\end{equation}
where the signal strength is given by
\begin{equation}
S^2 = \int_0^T s(t)^2\, dt
\end{equation}
and
\begin{eqnarray}
\gamma_{ij} &=& \Bigl(
\int_0^T \Bigl[ \partial_i C_+\partial_j C_\times + 
C_\times \left(\partial_i C_+ \partial_j \Psi + \partial_j C_+
\partial_i \Psi\right) \nonumber \\
&& \quad + \partial_j C_+\partial_i C_\times
- C_+ \left(\partial_i C_\times \partial_j \Psi + \partial_j C_\times
\partial_i \Psi\right) \nonumber \\
&& \quad + \left( C_+^2+C_\times^2\right)\partial_i\Psi \partial_j\Psi
\Bigr] dt \Bigr) / \Bigl( \int_0^T \left( C_+^2+C_\times^2\right) dt \Bigr).\nonumber \\
\end{eqnarray}
Here $C_+(t) = F^+(t) A_+$ and $C_\times(t) = F^\times(t) A_\times$. The amplitude modulation
functions $C_+(t)$ and $C_\times(t)$ depend on $(\ln {\cal A}, \imath, \psi, \theta, \phi)$
but are independent of $f$ and $\varphi_0$, while the phase function $\Psi(t)$
depends on $(f, \theta, \phi, \varphi_0)$ but is independent of $\ln {\cal A}, \imath$ and
$\psi$. In other words, the amplitude modulation
terms are solely responsible for determining the intrinsic amplitude,
inclination and polarization, while the phase term is solely responsible for
determining the frequency and orbital phase. Both the amplitude and phase play
a role in fixing the sky location $(\theta,\phi)$.

\begin{figure}[ht]
\vspace{55mm}
\includegraphics{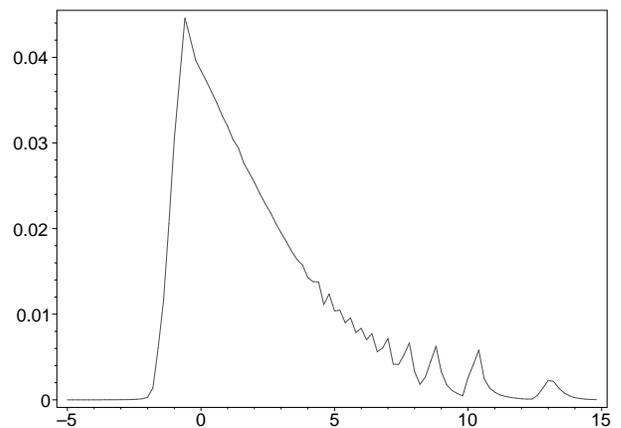}
\vspace{5mm}
\caption{A histogram of the quantity $\log_{10}({\rm vol}(\vec{\lambda}))$.}
\label{fig1}
\end{figure}

Using the expressions derived in Ref.~\cite{cr1} it is a simple, yet laborious task to
compute $\gamma_{ij}(f,\theta,\phi, \imath, \psi)$. The final expressions are long and rather
opaque. Most striking is the strong dependence on $(f,\theta,\phi, \imath, \psi)$, which
makes it exceedingly difficult to give simple answers to questions such as ``what
is the angular resolution of the LISA detector?''. One trend that is obvious is the improvement in
the information encoding at high frequencies that comes from the Doppler modulation. Thus, to
obtain a lower limit on the bandpass required to remove a source from the
data stream one should work in the very low frequency limit (which is equivalent to setting
$\Phi_D(t)=0$). Even then $\gamma_{ij}$ shows a strong dependence on
$(\theta,\phi, \imath, \psi)$ that defies efforts to define ``typical'' parameter
sensitivities. Using (\ref{vol}) and (\ref{fish}) we have
\begin{equation}\label{volx}
\Delta V(\vec{\lambda})= \left(\frac{N}{S}\right)^7 {\rm vol}(\vec{\lambda})\, ,
\end{equation}
where
\begin{equation}
{\rm vol}(\vec{\lambda})= \frac{\pi^3\sqrt{2}}{105}
\frac{1}{\sqrt{{\rm det}(\gamma_{ij}(\vec{\lambda}))}} \, .
\end{equation}
When calculating $\gamma_{ij}$ the dimensionless quantity $q= fT$ is
used instead of $f$. Physically, $q$ is the number of wave cycles in the observation
period. In terms of the coordinates $(q,\ln {\cal A}, \theta, \phi, \imath, \psi, \varphi_0)$,
the volume $V(B)$ is equal to
\begin{equation}\label{vb}
V(B) = 4 \pi^3 BT (\Delta \ln{\cal A}) \simeq  285 N_b \, 
\end{equation}
where the final expression uses the fact that ${\cal A}$ is typically within a factor of
$\sim 10$ of the fixed signal strength $S$, and $N_b= BT$ is the number of frequency bins
covered by the bandwidth $B$. Combing equations (\ref{sinfo}), (\ref{volx}) and (\ref{vb})
yields
\begin{equation}
{\cal I}(B,\vec{\lambda}) = 7\log_2\left(\frac{S}{N}\right)+\log_2(285 N_b)
-\log_2({\rm vol}(\vec{\lambda})).
\end{equation}
While the quantity ${\cal I}(B,\vec{\lambda})$
does vary with the binary's location in parameter space $\vec{\lambda}$, the variations are
tamed by the logarithmic dependence on ${\rm vol}(\vec{\lambda})$. Figure~\ref{fig1} shows
a histogram of the quantity $\log_{10}({\rm vol}(\vec{\lambda})$ in the very low frequency
limit ($f \sim 0.1$ mHz) for $T={\rm 1 yr}$ of observations.
The histogram was generated by sampling $(\theta,\phi)$
on a uniform HEALPIX grid~\cite{HEALpix} with 49152 pixels, and by sampling both $\imath$
and $\psi$ uniformly at 200 points, making for a grand total of $1.97 \times 10^9$
samples. The quantity $\log_{10}({\rm vol}(\vec{\lambda}))$ was found to have a median value
of $1.4$ and a mean value of $2.3$. Thus, the average value of ${\cal I}(B,\vec{\lambda})$
is equal to
\begin{equation}\label{Iav}
{\cal I}(B) = 7\log_2\left(\frac{S}{N}\right)+\log_2(1.43\, N_b) \, .
\end{equation}
Solving the transcendental equation (\ref{tran}) yields
\begin{equation}
N_b \simeq 7 + \frac{\log_2(10)}{\log_2(S/N)} .
\end{equation}
Thus, for typical sources, $N_b \simeq 8$. This result is in striking disagreement with the
recent work of Kr\'{o}lack and Tinto~\cite{massimo}, where a Fisher matrix based approach was
used to derive the result $N_b \simeq 0.5$. They argue that the number of sources than can
be removed scales as $V(B)/\Delta V(\vec{\lambda})$, which is precisely opposite to the
reasoning used to derive Eqn.~(\ref{sinfo}).

It has been estimated~\cite{gils} that the number of detached wd-wd binaries
per $1/{\rm yr}$ frequency bin scales as
\begin{equation}
N_{\rm bin} = \left(\frac{f}{1.445\,{\rm mHz}}\right)^{-11/3} \, .
\end{equation}
Setting $N_{\rm bin}= 1/8$ yields $f=2.54$ mHz for the frequency cutoff below which
the galactic population of wd-wd binaries becomes a confusion noise source for LISA.
Increasing the limit from $1.445$ mHz to $2.54$ mHz may not sound like much, but it translates
into a significant reduction in the number of galactic binaries LISA can resolve.

The ``eight bin rule'' derived above is likely a fairly robust estimate. Mostly it comes
from the overall $(N/S)^7$ factor that scales the size of the seven dimensional error
ellipsoid. The estimate is fairly insensitive to how $\gamma_{ij}$ is calculated, so
including the Doppler modulation will only slightly increase $N_b$. As argued earlier,
utilizing the full stereo capabilities of LISA would reduce $N_b$, but by less than a
factor of two. In Ref.~\cite{cl} a single channel was used to test an algorithm for
identifying and subtracting binary systems from the LISA data stream. It was found that
the subtraction worked for 3 sources separated by an average of 5 bins. This appears to
contradict the eight bin rule, but information from surrounding bins was also used. The
number of bins with $S/N > 5$ that contributed to the subtraction totaled $\sim 40$, so
approximately 13 bins were used to remove each signal, and even then the subtraction
was imperfect. One of the key questions to be addressed in future work is how close can
one get to the information theory limit. At least now we have a better idea of
what that limit is.

\section*{Acknowledgments}

This work grew out of discussion with Sterl Phinney, Ron Hellings and Shane Larson. Financial
support was provided by the NASA EPSCoR program through Cooperative Agreement NCC5-579.

\end{document}